\documentclass[amsmath,10pt]{revtex4}

\usepackage{graphicx}
\usepackage{tikz}

\usepackage{lipsum}
\usepackage{ulem}
\usepackage{amsmath}

\begin{document}

\title{Coherent Perfect Absorption induced by the nonlinearity of a Helmholtz resonator} 
\author{V. Achilleos, O. Richoux and G. Theocharis}
\affiliation{LUNAM Universit\'{e}, Universit\'{e} du Maine, CNRS, LAUM UMR 6613, avenue O. Messiaen, 72085 Le Mans, France.}
\maketitle

\section*{Abstract}

In this work, we analytically report Coherent Perfect Absorption induced by the acoustic nonlinear response of a Helmholtz Resonator 
side loaded to a waveguide. In particular, we show that this two-port acoustic system can perfectly absorb two high amplitude symmetric incident waves when the additive nonlinear losses in the HR, induced by the jet flow separation, together with the weak linear viscothermal losses of the HR balance the radiation losses to the waveguide. For the case of the one-sided incidence configuration, this condition leads to an absorption equal to $0.5$.
This result, which is verified experimentally, is in a good agreement with an analytical nonlinear model of the impedance of the HR. The nonlinear control of perfect absorption using resonators will open new possibilities in the design of high amplitude sound attenuators for aero-engine applications. 

\section{Introduction}

The inherent viscothermal losses of an acoustic system combined with wave interference can give rise to
perfect absorption, a phenomenon which is of great applied interest 
in many fields like room acoustics \cite{Kuttruff,Cox}, duct mufflers \cite{Munjal}, aeroacoustic liners \cite{Munjal} and environmental acoustics \cite{Kotzen}. 
In the particular case of a two-port, resonator/waveguide system, it has been shown that Coherent Perfect Absorption (CPA) can be obtained under a so-called critical coupling condition 
when the two-sided incident waves correspond to an eigenmode of the scattering matrix of the system. The critical coupling condition is achieved when the leakage rate of energy out of the resonant system (resonant system-waveguide coupling) is balanced by the resonator losses. 
This phenomenon has been extensively studied in several fields of wave physics and especially in optics \cite{Haus,PA1,PA2}. 

In acoustics, the improvement of low frequency sound absorption constitutes a real scientific challenge for the major issue of noise reduction. Resonant structures based on Helmholtz Resonators (HR), bubbles and membranes (sub-wavelength scatterers) provide excellent candidates for the design of efficient, thin and light absorbing structures. 

Typically, aforementioned resonators exhibit weak inherent losses. Thus, 
for a single resonator/waveguide system, critical coupling condition can be fulfilled by using either a highly lossy resonator, for 
example a poroelastic membrane \cite{Romero_Report} or viscous metascreens including bubbles \cite{bubbles}, or by increasing 
artificially the losses in the resonator by filling part of the cavity of the HR with porous material \cite{Romero_JASA}. On the other 
hand, when the system is composed of several resonators, the interaction of two or more resonant modes of moderate Q-factor can be 
also used to achieve the critical coupling of the structure \cite{Merkel}. 
Recently, impedance matched membranes have been also used to turn acoustic reflectors into perfect absorbers \cite{Ping}.

In this work we propose an alternative way to achieve the critical condition and thus CPA, using  nonlinear effects
in a HR to increase its weak linear losses and balance the leakage.

In the case of a HR coupled to a 1D waveguide, it is known that  sufficiently large amplitude  incident waves
lead to increased absorption, due to the conversion of acoustical energy  into kinetic rotational energy at the edges of
the resonators neck~\cite{sugi1,chinese,singh1,boek}.
Note that other nonlinealirities as for example a cubic Kerr-type nonlinearity, can be employed to control the CPA as a function of
the incident field amplitude~\cite{NCPAO}.
Control of absorption by the wave amplitude with sub-wavelength structures constitutes a topic of great theoretical interest and can find applications in noise reduction, and in the absorption of high amplitude pressure waves in aircraft engines or rocket launch pads. 
By means of a simplified nonlinear model of the HR, we obtain the CPA conditions and experimentally 
validate a maximum absorption equal to 0.5 at the critical coupling, for an one-sided incidence configuration. To further understand the underlying physical mechanisms, experimental results are compared to analytical ones based on a nonlinear impedance model showing a good agreement.

\section{Theory}
\begin{figure}[tbp!]
\includegraphics[width=14cm]{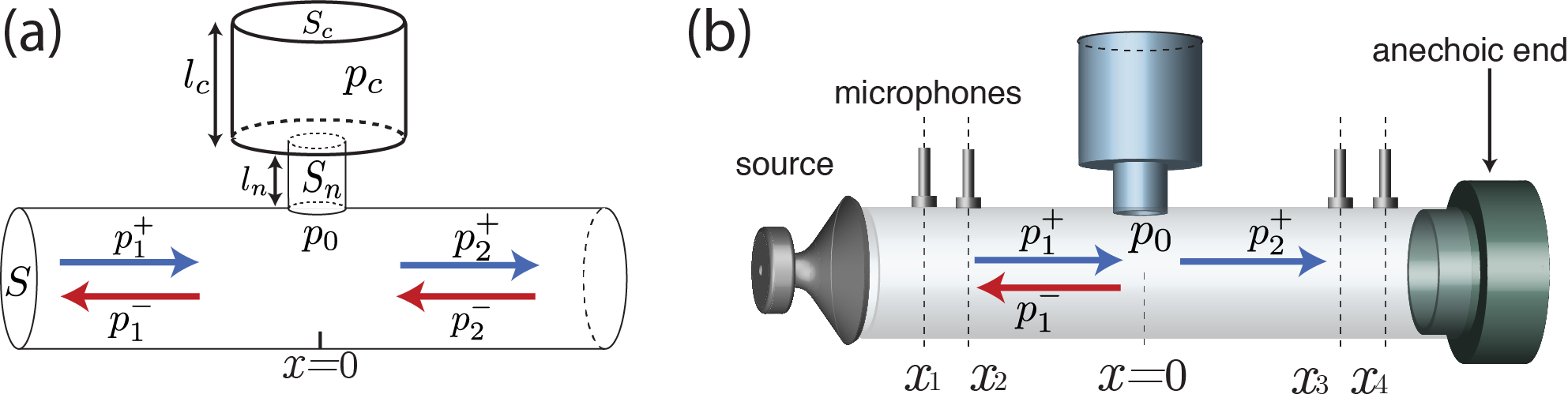}
\caption{ (a) Schematic representation of the system under study: a Helmholtz resonator side-loaded to a cylindrical waveguide.
The two-port scattering process is also indicated by the arrows. (b) The experimental setup used for our measurements.}
\label{theor} 
\end{figure}

Our system is composed of a HR side-loaded at $x=0$ to a waveguide, see Fig.~\ref{theor}. In this section, we derive simplified 
equations that describe the dynamics of the system in the general case of two, high pressure incident waves.
The HR that we consider is composed by a cylindrical neck, with a cross-section  $S_n$ and length $l_n$, and a
cylindrical cavity of cross-section $S_c$ and length $l_c$.  For the low frequency range, the related wavelengths $\lambda=2\pi/k$ are much larger 
to the geometric characteristics of  the neck, $k l_n\ll 1$, and of the cavity, $k(S_cl_c)^{1/3}\ll 1$. The former leads to the neglection 
of the compressibility of the fluid in the neck
while the later to the assumption that the pressure field inside the cavity is uniformly distributed. 

In this case, the dynamics of the pressure in the cavity of the resonator $p_c(t)$, can be described by the following approximate equation~\cite{sugi1,chinese,singh1,boek} :
\begin{eqnarray}
\ddot{p}_c+\omega_0^2p_c+R_L\dot{p}_c+\alpha|\dot{p}_c|\dot{p}_c=\omega_0^2p_0,
\label{ode1}
\end{eqnarray}
where ($\dot{}$) indicates differentiation with respect to time, $p_0$ denotes the pressure at the entrance of the resonator (see Fig.~
\ref{theor}),
$\omega_0^2=c_0^2S_n/l'_nl_cS_c$ is the resonance frequency of the HR, and $c_0$ is the 
speed of sound.  $l_n'$ indicates the corrected neck length $l_n'=l_n+\delta l$~\cite{Kergomard1,Dubos} taking into account both the acoustic radiation into the cavity and the waveguide.
The resistance factor  $R_L$ is a small parameter which quantifies the viscothermal losses in the resonator~\cite{singh1,Melling}.
The parameter $\alpha^{-1}$ of the nonlinear term in Eq.~(\ref{ode1}), depicts a characteristic pressure beyond which nonlinear losses become important. 
It is connected with flow separation and vorticity in the neck and it is approximated by $\alpha^{-1}=2C_{vc}^2\rho_0\omega_0^2 
l_n^{'2}$ (see~\cite{sugi1,singh1,boek}) where $\rho$ is the density of air and $C_{vc}^2$ is the {\it vena-contracta} coefficient
having a value of $\approx 0.7$ for a neck with hard edges.
The nonlinear response of the HR presents different behavior depending on the detailed geometrical characteristics of the edges of 
the neck~\cite{Ingard,Dissel,yve,temiz}. Different regimes of this behavior  can be quantified by the Strouhal 
number $\mbox{St} = \omega d_n/|u_n|$  where $u_n$ is the particle velocity in the neck of the resonator. 
When $\mbox{St}\gg 1$, particle displacement is smaller than the diameter of the neck and the system is linear. 
 In the opposite, strongly nonlinear limit, $\mbox{St}\ll 1$ and particle velocities are large enough so that vortices 
 are created and are blown away, as a jet, out of the resonator neck. The dynamics, 
described by  Eq.~(\ref{ode1}), is a good approximation for these limiting cases, and in the strongly nonlinear regime, the term analogous to $\alpha^{-1}$, quantifies the acoustic energy jetting out of the resonator. On the other hand, for moderate values
of $St$ and especially for $St\approx 1$,  there is no analytical model to describe the nonlinear behavior of the HR but, for
example, correction functions were proposed in Ref.~\cite{temiz} to provide a link between the two limits.

The waveguide, in which the HR is side-loaded, is a cylindrical duct of cross-section $S$. 
For a frequency range well below the first cut-off frequency of the waveguide, 
propagation  is considered one-dimensional. In addition, for small lengths of the waveguide, to avoid the accumulative
nonlinear behavior, it can be considered linear. Thus, the acoustic  propagation is described by the following
linearized mass and momentum conservation laws:
\begin{eqnarray}
\frac{\partial p(x,t)}{\partial t}+c^2\rho_0\frac{\partial u(x,t) }{\partial x}=0 ,\quad \rho_0\frac{\partial u(x,t)}{\partial t}+\frac{\partial p(x,t) }{\partial x}=0 \label{mm} 
\end{eqnarray}
where $p(x,t)$ and $u(x,t)$ are the pressure and particle velocity in the waveguide. For a HR located at $x=0$, we can write the 
acoustic field inside  the waveguide as:
\begin{align}
&p(x,t)=\Bigg\lbrace \begin{array}{l}
p_1^+(\xi_+)+p_1^-(\xi_-) \quad x\leq 0,\\
{}\\
p_2^+(\xi_+)+p_2^-(\xi_-) \quad x\geq 0,
\end{array} \label{ansatz}
\end{align}
where  $\xi_\pm=t\mp x/c_0$, while $^+$ and $^-$ denote right- and left-going waves respectively. Considering the 
HR as a point scatterer, continuity of pressure 
 at $x=0$  yields the relation:
\begin{eqnarray}
p_1^+(t)+p_1^-(t)=p_2^+(t)+p_2^-(t)=p(0,t)\equiv p_0.\label{cont}
\end{eqnarray}
The linearized conservation of mass in the HR cavity is given by  
\begin{eqnarray}
u_n=\frac{S_cl_c}{c_0^2\rho_0S_n}\dot{p}_c. \label{massc}
\end{eqnarray}
Additionally, using conservation of acoustic flux at $x=0$ in the waveguide,
combined with Eqs.~ (\ref{mm}), (\ref{ansatz}) and (\ref{massc})
we find the following relation
\begin{eqnarray}
p_1^{'+}\Bigr|_{x=0}-p_1^{'-}\Bigr|_{x=0}=p_2^{'+}\Bigr|_{x=0}-p_2^{'-}\Bigr|_{x=0}+\frac{S_cl_c }{c_0S}\ddot{p}_c, \label{disc2}
 \end{eqnarray}
where the primes denote differentiation with respect to $\xi_{\pm}$. By integrating Eq.~(\ref{disc2}) in time, we obtain
\begin{eqnarray}
p_1^+(t,0)-p_1^-(t,0)=p_2^+(t,0)-p_2^-(t,0)+\frac{2\gamma}{\omega_0^2}\dot{p}_c,\label{disc3}
 \end{eqnarray}
where $\gamma=c_0S_n/2l_n'S$ is a coefficient  describing the coupling strength between the HR and the waveguide.
Finally, using Eqs.~(\ref{ode1}), (\ref{cont}) and (\ref{disc3}), the dynamics of the pressure $p_c(t)$, inside the HR cavity, is described by the following equation:
\begin{eqnarray}
\ddot{p}_c+\omega_0^2p_c+(R_L+\gamma+\alpha|\dot{p}_c|)\dot{p}_c=\omega_0^2(p_1^++p_2^-).
\label{ode3}
\end{eqnarray}
The incident waves from the left, $p_1^+$, and from the right, $p_2^-$, act as drivers. The additional dissipative term proportional 
to $\gamma$, describes the leakage rate of energy out of the HR (resonator-waveguide coupling). 
When the  ``boundary conditions'' (i.e. conditions for  incident or outgoing waves) are known, the 
scattering properties of the system are determined by Eqs.~(\ref{cont}), (\ref{disc3}) and (\ref{ode3}).
Below we focus on two examples of such conditions i) a case with  two-sided incidence and ii)
a case of an one-sided incidence combined with an anechoic termination.

\subsection{Two-sided incidence and CPA}

In the following,  we investigate the boundary condition which corresponds to the phenomenon of CPA
for monochromatic incoming waves, i.e.  $p_1^+$, $p_2^-\propto\exp[-i \omega t]$.
For the two-sided incident system, CPA boundary condition implies  that outgoing waves must  vanish, namely  $p_1^-=p_2^+=0$. 
From Eq.~(\ref{cont}),  this requirement implies $p_1^+=p_2^-$ which corresponds to a \textit{symmetric} CPA~\cite{APL}.
Moreover, from Eqs.~ (\ref{cont}) and (\ref{disc3}),  we obtain that $p_1^++p_2^-=\frac{2\gamma}{\omega_0^2}\dot{p}_c$ and
Eq.(\ref{ode3}) can be written as 
\begin{eqnarray}
\ddot{p}_c+\omega_0^2p_c+(R_L-\gamma+\alpha|\dot{p}_c|)\dot{p}_c= 0.
\label{ode4}
\end{eqnarray}
Note that due to the CPA condition, the coupling term $-\gamma \dot{p}_c$ in Eq. (\ref{ode4}), appears as  an effective gain.
Since we are interested in the frequency range close to the resonance, and for sufficiently small pressure amplitudes we consider
the nonlinear damping $\alpha|\dot{p}_c|\dot{p}_c$ to be of the same order as the linear term $(R_L-\gamma)\dot{p}_c$.
For the HR used in our experiments
and for cavity pressures of approximately $170$~dB the nonlinear coefficient $\alpha|\dot{p}_c|/\omega_0$ 
and the linear damping coefficient $ (R_L-\gamma)/\omega_0$ are of the same order $\epsilon$ and take values~$\approx 0.2$. 
Thus, similarly to \cite{singh1}, we look for the stationary solution of Eq.~(\ref{ode4}) applying the Lindstedt-Poincar\'e perturbation technique \cite{Nayfeh} up to first order, neglecting the contribution of higher harmonics. 
The requirement of the annihilation of the secular terms, leads to the conclusion that this particular type of nonlinearity does not introduce a frequency shift, in this order, and imposes the following condition
\begin{equation}
\frac{4}{3\pi}\frac{|u_n|}{l_n}+R_L=\gamma.
\label{result}
\end{equation}

Thus, CPA occurs at the resonance frequency  $\omega_0$, under symmetric and in-phase incoming waves, $p_1^+=p_2^-$ 
and when the linear $R_L$ and nonlinear $\frac{4}{3\pi}\frac{|u_n|}{l_n}$ decay rates are balanced with the leakage rate $\gamma$. 
Note that the CPA condition depends on the cross-section of the waveguide,  and on the geometry of the HR neck
on the cavity. As such, the same condition may apply for different frequencies by tuning the length of the cavity.

\subsection{One-sided incidence and critical coupling	}
Eq.~(\ref{result})  can be seen as an amplitude-dependent extension of the critical coupling condition~\cite{Haus,PA1,PA2}.
It was recently shown that, for the case of an incident wave from the left and an anechoic termination on the right 
of the waveguide in the linear regime,
a critically coupled HR  leads to  a maximum absorption of $0.5$~\cite{Merkel}. 
For the considered setup, we can define the transmittance  $T$, reflectance $R$ and absorption $A$ as
\begin{eqnarray}
T=\left|\frac{p_2^+}{p_1^+}\right|^2 ,\quad R=\left|\frac{p_1^-}{p_1^+}\right|^2
\quad A=1-T-R.
\label{TnR}
\end{eqnarray}
The absorption $A$, when the critical coupling condition Eq.~(\ref{result}) is satisfied, can be calculated from
Eqs.~(\ref{cont}), (\ref{disc2}) and (\ref{ode3}), and it is also found to be $A=0.5$, for the case of a HR with nonlinear losses. 
\section{Experimental results}
We now turn to one-sided incidence experiments to confirm the
nonlinear critical coupling. To do so, we measure the absorption for different amplitudes and frequencies of incident waves $p_1^+$ and verify that nonlinear losses can lead to an absorption of $A=0.5$ when condition~(\ref{result}) is satisfied.
\subsection{Setup}
We perform experiments using the apparatus shown in the right panel of  Fig.~\ref{theor}. The 
geometrical characteristics of the setup are: 
$S=20 \times 10^{-4}$~m$^2$, $S_n=3.14 \times 10^{-4}$~m$^2$,
 $S_c=14.5\times 10^{-4}$~m$^2$, $l_n=0.02$~m and $l_c=0.038$~m. 
A compression driver (Beyma CP800TI), tuned to provide high amplitude waves with very low harmonic distortion, connected to a 
linear amplifier (Devialet D-Premier), is used as a source at one side of the $50$ cm long waveguide.  An anechoic termination is 
located at the 
other side providing reflection below $5\%$ for the considered range of frequencies and amplitudes.
Measurements of the pressure are performed at four different positions $x_i$ using a
$1/4$~inch B\&K free field microphone. Assuming that the propagation is linear in the waveguide, we experimentally determine $T$, $R$ and $A$ by using the four-microphones method \cite{Song,Muehleisen}.
This procedure is repeated  for several values of the amplitude of the incident wave $|p_1^+|$ from approximately $105$ dB to $160$ dB.
\subsection{Nonlinear impedance model}
To compare the experimental results with the theory of the previous section, we note that a nonlinear 
impedance of the resonator can be obtained from Eq.~(\ref{ode1}), and it can be written as~\cite{singh1}
$
Z_{HR}= i\mathcal{X}+\mathcal{R}_L + \mathcal{R}_{NL}(\hat{u}_n), \label{zhr} 
$
where $\mathcal{X}$, $\mathcal{R}_L$ and $\mathcal{R}_{NL}(\hat{u}_n)$ are the linear 
reactance, linear resistance and the nonlinear resistance respectively. The linear part of the impedance is found to be $i\mathcal{X}+
\mathcal{R}_L=\rho_0 l_n\big(i(\frac{\omega_0^2-\omega^2}{\omega})+R_L\big)$ while the nonlinear:
\begin{eqnarray}
\mathcal{R}_{NL}(\hat{u}_n)=\frac{4\rho_0|u_n|}{3\pi S_nC_{vc}^2}.,
\label{nlnZ}
\end{eqnarray}
We note that the linear part, which is obtained from the simplified model of Eq.~(\ref{ode1}), is valid for wavelengths sufficiently 
larger than the geometric characteristics of the neck and of the cavity. The HR used in our experiments is not strictly in this regime. 
A more accurate expression of the linear HR impedance can be obtained by the transfer matrix method, see
appendix of~\cite{Merkel} which predicts a resonance frequency of $f_0'=710$~Hz instead of $f_0=\omega_0/2\pi=756$~Hz.
The nonlinear resistance in Eq.(\ref{nlnZ}) is linearly dependent on the velocity amplitude, and it is known to be a good
approximation in the strongly  nonlinear regime ($St\ll 1$)~\cite{Ingard,temiz}. Nevertheless, as it is mentioned in Section II, 
 the nonlinear response of the resonator highly depends on the detailed geometry of the neck edges, the Strouhal number 
 and other factors such as the shear number.
Thus, in order to have a better agreement with the experimental results, we use a nonlinear resistance $\mathcal{R}_{NL}\rightarrow F_c \mathcal{R}_{NL}$, where  $F_c$ has the following form
\begin{eqnarray}
Fc=\frac{1}{1+\Delta St},
\label{fit}
\end{eqnarray}
while the parameter $\Delta$ is a fitting parameter. The function $F_c$ is chosen according to 
the recent results of Ref.~\cite{temiz} to satisfy the corresponding known limits of linear ($\mbox{St}\gg1$) and 
nonlinear ($\mbox{St} \ll 1$) response of the resonator.
Using the definitions in Eq.~(\ref{TnR}) ,we can write the transmittance and reflectance as functions of the impedance 
\begin{eqnarray}
T=\left|\frac{1}{1+\sigma}\right|^2,\quad R=\left |\frac{\sigma}{ 1+\sigma}\right|^2,\label{TnR2}
\end{eqnarray}
where $\sigma=c_0S_n\rho_0/2SZ_{HR}$.
%
%

\subsection{Results}
\begin{figure}[h!]
\includegraphics[width=14cm]{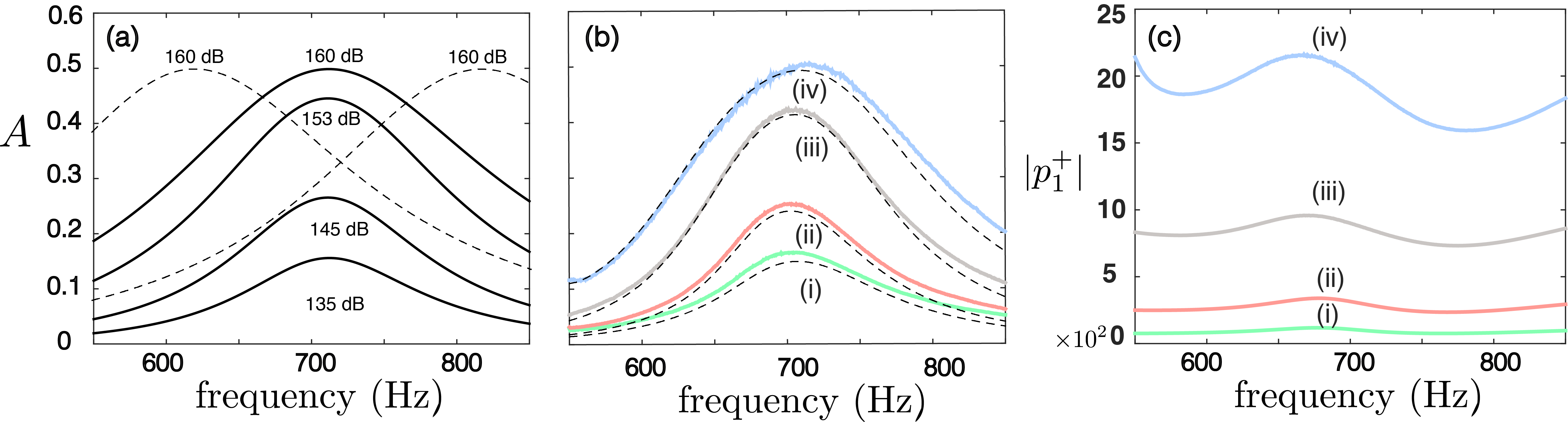}
\caption{(Color online) (a) Absorption as a function of frequency for different incident amplitudes   $|p_1^+|$ as
obtained using the nonlinear impedance with the fitting parameter $\Delta=0.15$. Left (right) shifted, dashed line
corresponds to the case of a HR with cavity length $l_c=5$~cm ($l_c=3$ ~cm). (b) Solid lines depict the
measured absorption for different amplitudes of incident waves  $|p_1^+|$  represented in panel (c).
Dashed lines in panel (b) show the theoretical prediction obtained by the  nonlinear impedance with $\Delta=0.15$.}
\label{figtheor} 
\end{figure}

In  panel (a) of Fig.~\ref{figtheor}, we plot the theoretically obtained absorption for different values of the incident
amplitude $\vert p_1^+\vert$,  using the nonlinear impedance model with $\Delta=0.15$, for a HR corresponding to our experimental setup. As it is observed, for increasing incident 
amplitudes the maximum of absorption increases, and reaches the value of $A=0.5$ for  $\vert p_1^+\vert=160$~dB at the resonance 
frequency $f_0'$. The dashed lines correspond to the same amplitude of $\vert p_1^+\vert=160$~dB but for different cavity lengths, 
$l_c=5$~cm (left shifted line) and $l_c=3$ ~cm (right shifted line), verifying that $A=0.5$ can be reached at the same amplitude, for 
different frequencies by tuning the length of the cavity. In panel (b) of Fig.~\ref{figtheor}, the solid 
lines depict the experimentally obtained absorption with the corresponding experimental incident wave amplitudes shown 
in panel (c). Note that small variations of the frequency of  maximum absorption observed in panel (b), are due to the fact that 
the incident wave amplitudes are  frequency dependent caused by the coupling between the reflected wave and the transducer.
The dashed lines in panel (b), show the theoretical prediction as obtained using the nonlinear impedance and the fitting function 
$F_c$ with $\Delta=0.15$, where the velocity of the neck is calculated using the experimental value s of the incident 
pressure shown in panel (c). The good agreement between theory and experiment demonstrates that the nonlinear response of the HR 
is adequately described by the nonlinear impedance with the additional fitting.

\begin{figure}[h!]
\centering
\includegraphics[width=7cm]{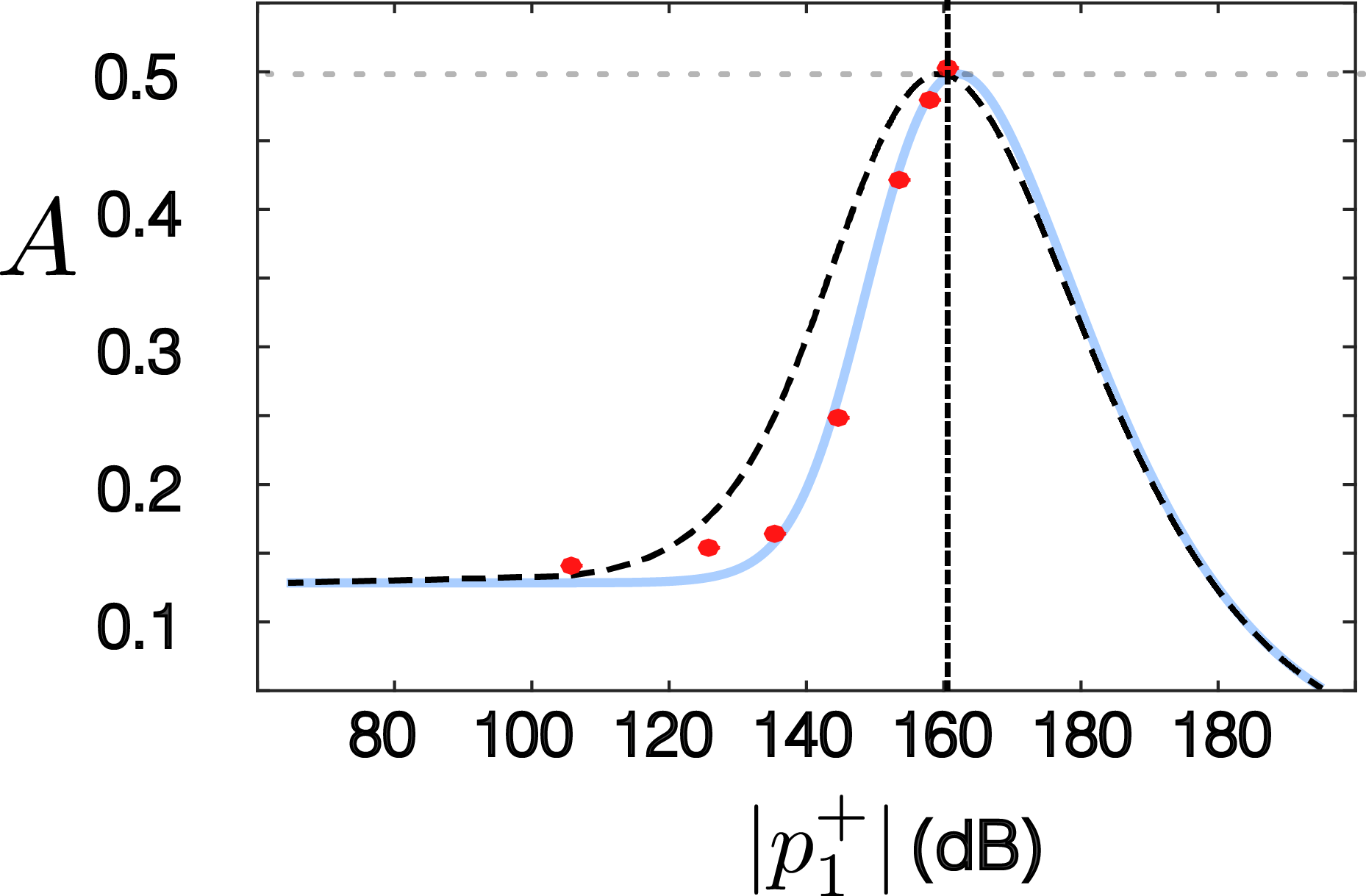}
\caption{(Color online) (a) Absorption at the frequency $f_0'=710$~Hz as a function of 
the amplitude of the incident wave  $|p_1^+|$. Blue solid (dashed) line corresponds to the nonlinear impedance model 
using the fitting parameter $\Delta=0.15$ ($\Delta=0$). Dots indicate the corresponding experimental 
measurements. The vertical dashed line depicts the prediction of Eq.~(\ref{result}) for the CPA condition. 
}
\label{expgood} 
\end{figure}

To directly compare the experimental results with the analytical condition for CPA given by  Eq.~(\ref{result}), in Fig.~\ref{expgood}
we show the absorption for the resonance frequency $f_0'$, as a function of  $\vert p_1^+\vert$. 
The (red) dots depict the experimentally obtained values, and $A=0.5$ is reached for an incident amplitude of   $160$~dB.
More importantly, this is the value predicted by  Eq.~(\ref{result}) as indicated by the vertical dashed line, and
it verifies  that: critical coupling can be induced by nonlinear 
losses when, in addition with the linear ones, they balance the leakage to the waveguide. 
The blue solid line is obtained using the theoretical nonlinear HR impedance with $\Delta=0.15$, while
the dashed line corresponds to the case of $\Delta=0$ and thus to the prediction of the simplified model of Eq.~(\ref{ode1}).
Comparing the two theoretical curves, one observes that indeed at the two limits  (linear  $S_t \gg 1$ and strongly 
nonlinear $S_t \ll 1$), they coincide as expected (see discussion after Eq.~\ref{fit}). However in the intermediate 
regime (between $110$ dB and $160$ dB), the impedance with the fitting function describes more accurately the 
experimental results.


\section{Conclusions}
In this work we present a simplified model, to study the linear wave propagation in a waveguide, side-loaded with
a Helmholtz  resonator which exhibits nonlinear losses. Through this model, we extract the suitable conditions depending
on the geometrical characteristics and on the amplitude of the incident waves, in order to achieve CPA.
These conditions imply that CPA occurs at the resonance of the HR, when  the linear and the additive nonlinear losses of the resonator, balance the leakage to the waveguide under symmetric two-sided incidence.
For the one-sided incidence case, when the energy leakage criterion is satisfied, the HR is critically coupled giving rise to a maximum of absorption $A=0.5$. Extension of this study into more complex configurations could lead to the design of broadband high amplitude acoustic absorbers with applications into aeronautics.

We thank Y. Auregan, V. Pagneux and A. Hirschberg for helpful comments. This work has been funded by the Metaudible project ANR-13-BS09-0003, co-funded by ANR and FRAE.

\end{document}